\documentclass[twocolumn,prd,showpacs]{revtex4}
\usepackage{graphicx}
\usepackage{bm}
\usepackage{amsmath}
\usepackage{aas_macros}
\usepackage{subfigure}
\usepackage{epsfig}
\usepackage{footnote}

\newcommand{\lsim}{\raisebox{-0.3ex}{\mbox{$\stackrel{<}{_\sim} \,$}}}
\def\da {$d_{\rm A}(z)$ }
\def\dn {$\frac{dN}{dz}$ }
\def\mlim {$M_{\rm{lim}}$ }

\begin{document}
\preprint{}

\title{Using clusters in SZE + X-Ray surveys\\ as an ensemble of rulers to
constrain cosmology}

\author{Satej Khedekar and Subhabrata Majumdar}
\email{satejk@tifr.res.in, subha@tifr.res.in}
\affiliation{Tata Institute of Fundamental Research, Homi Bhabha Road, Colaba,
Mumbai - 400076, India.}

\begin{abstract}
Ongoing and upcoming surveys in x-rays and SZE are expected to jointly detect
many clusters due to the large overlap in sky coverage. We show that, these
clusters can be used as an ensemble of rulers to estimate the angular diameter
distance, \da. This comes at no extra observational cost, as these
clusters form a subset of a much larger sample, assembled to build cluster
number counts \dn. On using this \da, the dark energy constraints can be improved by
factors of 1.5 - 4, over those from just \dn. Even in the presence of a mass
follow-up of 100 clusters (done for mass calibration), the dark energy
constraints can be further tightened by factors of 2 - 3 . Adding \da from
clusters is similar to adding $d_{\rm L}(z)$, from the SNe observations; for eg., \dn
(from ACT/SPT) plus \da is comparable to \dn plus $d_{\rm L}(z)$ in constraining
$\Omega_m$ and $\sigma_8$.

\end{abstract}

\pacs{95.35.+x, 98.80.-k, 98.65.Cw, 98.62.Py}

\maketitle

\section{Introduction}
Large cluster surveys like the SPT, ACT, Planck and eROSITA promise to
detect from a thousand to a few hundred thousand clusters in the
 coming decade. The abundance and redshift distribution
$\frac{dN}{dz}$ of these clusters are important probes to
understand the nature of dark energy as well as to
constrain other cosmological parameters like $\Omega_m$ and $\sigma_8$
\citep{Holder01, WangSteinhardt, Levine, Weller02, Hu}. To deduce a cosmology
from these $\frac{dN}{dz}$ observations, one requires a precise knowledge of the
limiting mass of the survey as a function of redshift. One frequently uses proxy 
observables such as X-Ray surface brightness and temperature \citep{Ebeling00}, Sunyaev
Zel'dovich effect (SZE) decrement \citep{Staniszewski, Hinks}, cluster richness
\citep{Postman, Koester} and lensing \citep{Wittman06, Zitrin09} for the masses
 of clusters, related through simple power-law scaling relations. These scaling
parameters are highly degenerate with the cosmological parameters, and breaking 
this degeneracy is crucial to obtain tight constraints on cosmology. This may realized,
for example, through the so called `self-calibration' techniques  \citep{MM04,
Hu, LimaHu04}. Other approaches include an `unbiased' mass follow-up of a
sub-sample of the survey clusters \citep{MM03,MM04} or better theoretical
modeling of clusters to predict the form of mass-observable scaling relation
\citep{YoungerHaiman, Reid06, CM09}.
One can also try to optimize the cluster surveys so as to get the best possible
survey yield \citep{SatejPRL, Battye}.

Measurement of the angular diameter distance, $d_{\rm A}$, at the redshift of
the cluster using a combination of SZE and X-Ray observations have been
routinely made over the last 30 years. The results have suffered, in the past, from various systematics and
 reliable estimates have only been achieved recently with analysis of
 statistically significant samples of galaxy clusters \citep{Reese02,Bonamente}.
These new observations have demonstrated the power of using clusters to measure
`$d_{\rm A}$ vs $z$' and use it to study the expansion history of our Universe
\citep{Molnar}. However, these recent progress has been done with targeted observations. Since, targeted
 observations are costly, this approach limits the size of the sample of $\lsim 100$.
 In this {\it letter}, we show that one can build up an `ensemble' of \da by picking
 a sub-sample of clusters discovered in both X-Ray and SZE surveys with overlapping
 area and redshift coverage. Since, the surveys are already geared
towards getting clusters for $\frac{dN}{dz}$, we get the \da without any {\it extra}
 targeted observations. Addition of \da to \dn helps in tightening cosmological constraints,
 especially on dark energy equation of state. This is not surprising, since using \da
 from clusters is akin to adding $d_{\rm L}(z)$ information from supernovae (SNe) observations. 

The rest of this paper is organized as follows. In \S~2, we briefly discuss
cluster number counts and the procedure to estimate \da from SZE + X-Ray observations; 
in \S~3 we describe the surveys, the choice of cosmological and cluster
models and also summarize our methodology;  in \S~4, we forecast constraints on
cosmological parameters; and finally, we conclude in \S~5.

\section{Preliminaries}
\begin{figure*}[ht]
\centering
\subfigure[]{
  \includegraphics[width=8.5cm]{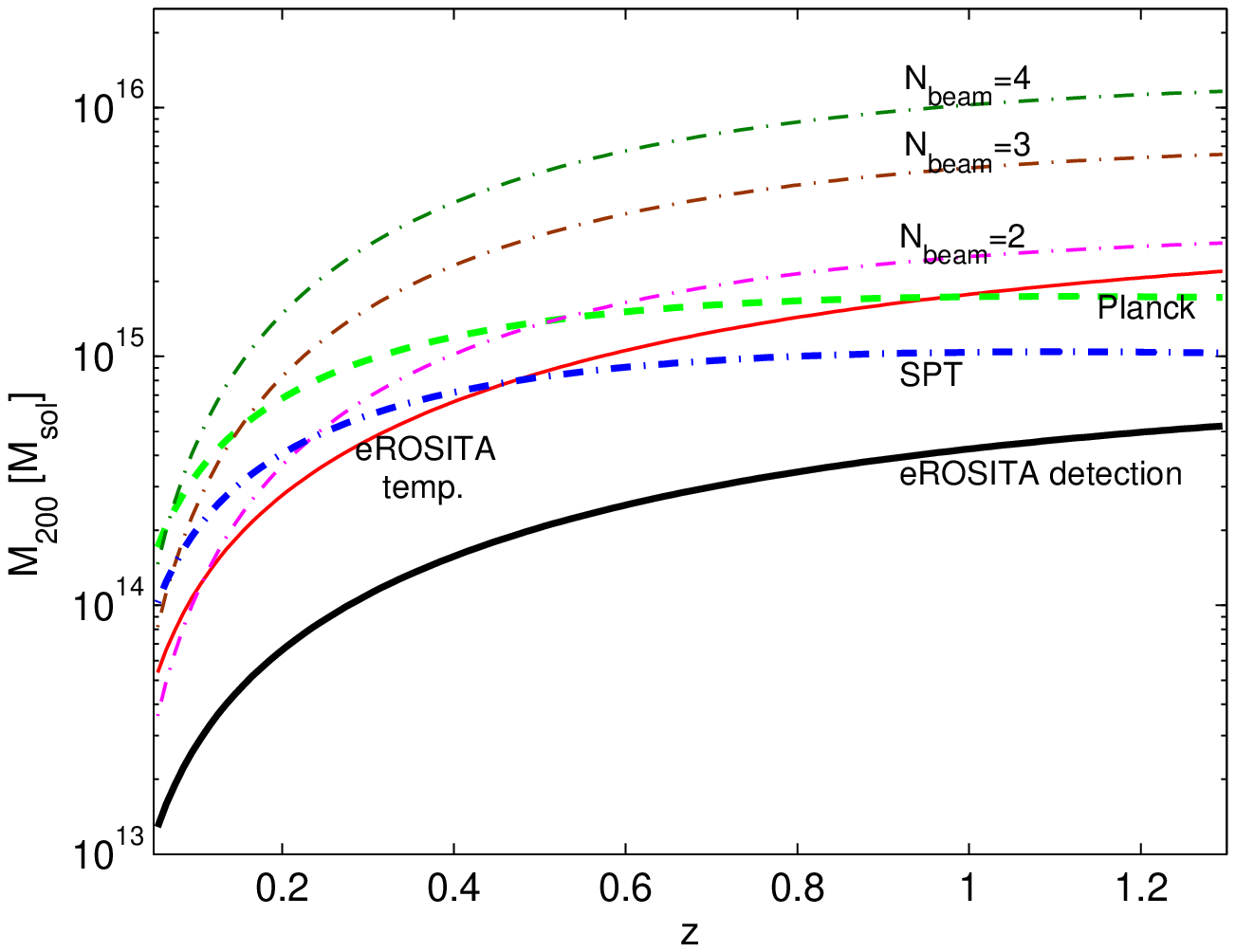}
  \label{fig:M_lim}
  \vspace{-4mm}
}
\subfigure[]{
  \includegraphics[width=8.5cm]{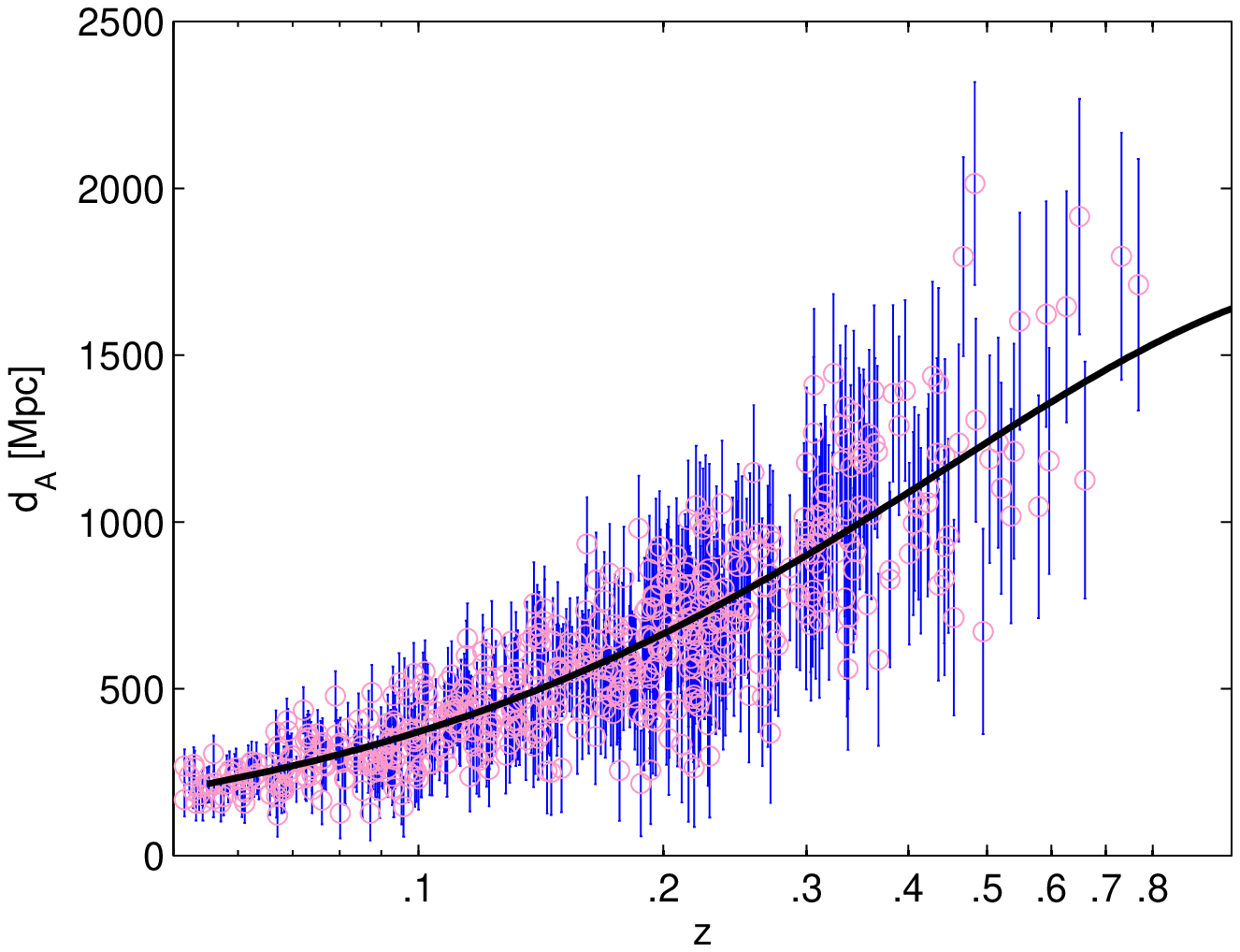}
  \label{fig:mock}
  \vspace{-4mm}
}
\label{fig:mocking}
\vspace{-4mm}
\caption[]{(a) $M_{\rm{lim}}$ as a function of redshift plotted
for various X-Ray and SZE surveys. The \da mock catalog is created by using the
effective $M_{\rm{lim}}(z)$, which is the maximum of the \mlim corresponding to a given $N_{\rm beam}$, the
 \mlim for determining temperature of the clusters detected through eROSITA and the \mlim for detecting clusters through Planck/SPT.
 (b) A mock \da catalog of 578 clusters (as found from an overlap of ACT/SPT with eROSITA)
 with 25\% Gaussian errors plus scatter. The thick line
indicates the $d_{\rm A}$ from fiducial cosmology.}
\end{figure*}

The redshift distribution of detectable clusters is given by,
\begin{equation}
\frac{dN}{dz}(z)=\Delta\Omega\frac{dV}{dzd\Omega}(z)\int_{0}^{\infty}\frac{dn(M,
z)}{dM}f(M,z)dM
\label{eqn:dndz}
\end{equation}
where $dV/dzd\Omega$ is the co-moving volume element, $f(M,z)$ is the
cluster selection function incorporating a logarithmic scatter
in the luminosity to mass conversion and $\frac{dn}{dM}$ is the halo mass
function taken from simulations \cite{Jenkins}.
For a X-Ray/SZE survey the limiting mass $M_{\rm{lim}}(z)$ is found from 
flux limit, $f_{\rm lim}$, of the survey. For X-Ray survey, we adopt 
luminosity-mass relations from \citet{ChandraVikhlinin} given by
$L_X=A_{X}\left(\frac{M_{500}}{10^{15}}\right)^{\alpha_{X}}E^{1.85}
(z)\left(\frac{h}{0.72}\right)^{-0.39}(1+z)^{
\gamma_{X}}$. For an SZE survey, our SZE flux-mass relation is
 $Y d_{\rm A}^{2}=A_{SZE} \left(\frac{M_{200}}{3.14 \times
10^{18}}\right)^{\alpha_{sz}}E^{2/3}(z)\left(1+z\right)^{\gamma_{sz}}$,
 where $Y$ is the integrated SZE distortion and \da is in Mpc.
 The parameters $\gamma_{X}/\gamma_{SZE}$ mimics any `non-standard'
evolution of cluster scaling relations \citep{MM03}. 
For the X-Ray survey, our fiducial parameter
values are: $\rm{log}(A_X) = -4.25$, $\alpha_X = 1.61$ and $\gamma_X = 0$ with
a  log-normal scatter of 0.246; while the corresponding values for the SZE
scaling relation are: $\rm{log}(A_{SZE}) = 1.75$, $\alpha_{SZE} = 1.61$, 
$\gamma_{SZE} = 0$ with a log-normal scatter of 0.2.

The distance $d_{\rm A}(z)$, of a cluster observed at a redshift
$z$ depends on the expansion history of the Universe as, $d_{\rm
A}(z)=\frac{c}{(1+z)}\int_{0}^{z}\frac{dz'}{H(z')}$
 where $H(z)$ is the Hubble expansion. The seminal paper by \cite{SilkWhite78}
defines the procedure to measure
\da from observations of clusters in SZE + X-Ray. This depends on `assuming' a
profile for the Intra-cluster medium, ICM, typically taken to be an isothermal
$\beta$-profile \footnote{Since the true ICM is neither
isothermal nor has $\beta$-profile, this leads to bias and scatter. The impact
is less for ensemble of clusters. Moreover,
the uncertainty can be calibrated to first order using cluster simulations.}
with $n_e(r) = n_{e_0} (1 + (r/r_c)^2 )^{-3\beta/2}$, where $n_e$ is the
electron number density, $r$ is the radius from the cluster's center, $r_c$ is
the core radius, and $\beta$ is a power-law index. Note, that one can always use a more
realistic ICM model with better data. For example, one can fit a temperature profile to improve upon the errors.
The X-Ray brightness, $S_X \propto \int n^2_e \Lambda_{ee} dl$,
 where $\Lambda_{ee}$ is the X-Ray cooling function of the gas; while for SZE,
$\Delta T \propto \int n_e T_e dl$, where $T_e$ is the ICM temperature.
Eliminating $n_e$, gives the angular diameter distance  $d_{\rm A} \propto
\frac{\Delta T^2_{\rm CMB} \Lambda_{ee}}{S_{X_0} T_e^2
\theta_c}$; see \citet{Birkinshaw99} for more details.

\section{Estimating constraints from future surveys}

\subsection{Fiducial cosmology, priors and survey descriptions}

We adopt our fiducial cosmology from the WMAP 7-year results (Table 6 of
\citep{WMAP7}) along with 
the following priors $(\Delta n_s, \Delta\Omega_b, \Delta h) = (0.015, 0.0037,
0.028)$.
For simplicity, we choose a flat Universe since for an open wCDM model, WMAP7+BAO+H0 tightly constraints 
$\Delta\Omega_{\rm tot} \leq 0.007$. With a small number of clusters, as suggested by recent SZ observations \cite{Vanderlinde}, 
just \dn data would not be sufficient to break the cosmology-cluster physics degeneracies.
 We therefore put priors on the scaling parameters $\Delta A = 0.003$ and
 $\Delta \alpha = 0.015$ motivated by recent observations \cite{Arnaud}.
 In addition, we put a weak prior of $\Delta \gamma = 0.2$ in all the cases.

We consider the following surveys with overlapping sky coverage - \\
(1) {\sf ACT/SPT}: We model the ongoing ACT/SPT survey as a $4000$ 
deg$^{\text{2}}$ with a $f_{\rm lim}$ of 75 mJy (at 150 GHz), so as to give $\lesssim$ 1000 clusters. \\
(2) {\sf Planck}: Ongoing all sky SZE survey. We take $f_{\rm lim}$ to be 300 mJy (at 353 GHz) which
returns $\lesssim$ 2000 clusters in $\sim 32000$ deg$^{\text{2}}$. The higher flux limit means that Planck would detect only 
massive low $z$ clusters.\\
(3) {\sf eROSITA}: Upcoming full sky X-Ray survey. We assume a [0.5-2.0 keV]
band  $f_{\rm lim}=4\times10^{-14}$ erg cm$^{\text{-2}}$ s$^{\text{-1}}$ which gives
us $\lesssim 1 \times 10^5$ clusters for $\sim 32000$ deg$^{\text{2}}$. All clusters detected by both Planck
 as well as ACT/SPT are expected to be detected by eROSITA as it has a much smaller $M_{\rm{lim}}$ (see Fig. \ref{fig:M_lim}).

\subsection{Methodology}
The redshift distribution of clusters is obtained by using
 the cluster scaling relations to convert flux to
 the corresponding lowest observable cluster mass $M_{\rm{lim}}(z)$. Figure
\ref{fig:M_lim} shows the $M_{\rm{lim}}$ for some of the surveys
 that we have considered. We further place a lower cut-off of $1.3 \times 10^{14}
h^{-1} M_{\odot} (M_{500})$ on $M_{\rm{lim}}$ and a redshift cut-off of 0.1 \footnote{This is done to get rid of
 the small mass nearby clusters which arise due to the steep nature of mass-flux
 relationship at very redshifts.}. The $\frac{dN}{dz}$ ($\Delta z$ = 0.1)
 likelihoods are computed using Cash C statistics.

We generate the mock $d_{\rm A}$ catalog as follows.
For a cluster to be visible in both X-Ray and SZE surveys, its mass must lie
 above the highest of the $M_{\rm{lim}}(z)$ from either of the surveys. To
estimate $d_{\rm A}$, one also needs {\it at least} a single isothermal temperature measurement of the ICM, 
 which is possible for 10 times the detectable flux; 
for this we calculate the corresponding higher $M_{\rm lim}$. Next, to fit a
$\beta$-model, we also require that
the clusters be well resolved; i.e. be larger than a certain minimum angular
size, so as to estimate
its core radius $\theta_c$. This is fulfilled by the condition that $\theta_c$
 be {\it at least $N_{\rm beam}$ times} the minimum resolution
 of the survey (16'' for eROSITA) \footnote{As practiced, the cluster structure
 fitting is done through X-Ray observations having better resolutions, while SZE observations mainly
 provide the central SZE decrement.}. We convert this to 
isophotal size $R_I=\theta_Id_{\rm A}$ and use the scaling relation
 between cluster mass and its size \citep{ohara_mohr} to enforce this
constraint. Fig. \ref{fig:M_lim} shows all the limiting masses
 described above as a function of redshift. The redshift $d_{\rm A}$ catalog is
constructed from [\ref{eqn:dndz}] by integrating over
 the highest of these $M_{\rm lim}(z)$'s in the plot. The \da from these clusters are
 distributed randomly with a Gaussian scatter of 25\% about the
 \da from fiducial cosmology (see Fig. \ref{fig:mock}). Such catalogs are created for
overlaps of SZE and X-Ray surveys like ACT/SPT + eROSITA and
Planck + eROSITA for $N_{\rm beam}$'s of 2, 3 and 4. A higher value of $N_{\rm beam}$ implies
a selection of only the larger clusters; see Fig. \ref{fig:M_lim}. The $d_{\rm
A}$ catalog is analyzed using a chi-square statistic for the likelihoods.

Finally, we do a joint analysis of the likelihoods from \da and those from number counts.
To forecast the constraints from cluster surveys we use MCMC
simulations in the parameter space of 6 cosmological
 parameters -- $\Omega_m$, $w_0$, $w_a$, $h$, $n_s$ and $\Omega_b$ and 3 scaling
parameters -- $A$, $\alpha$ and $\gamma$.

\begin{table*}[htb]\small
\caption{Comparison of $1-\sigma$ parameter constraints from ACT/SPT $\frac{dN}{dz}$ and its overlap with eROSITA.
 $N_{cl}$ is the number of clusters for which \da is measured and $N_{SNe}$ is the number of SNe. Columns 3 - 5
 lists improvements in the constraints obtained from ACT/SPT \dn when \da from various datasets are added. Column 6 shows
 the constrains when CMB priors are imposed on $\Omega_m$ and $\sigma_8$ from the WMAP7 results. The last two columns show
 the constraints when there are no priors on the scaling parameters $A$ and $\alpha$, but a mass follow-up of 97 clusters is
 added with (randomly distributed) errors on the masses of 15-50\% and 30-100\% for the follow-up's 1 and 2 respectively.\\}
\hbox to \hsize{\hfil\begin{tabular}{ccccccccc}
\hline
Parameter & \multicolumn{1}{c}{only $\frac{dN}{dz}$} &
\multicolumn{1}{c}{$\frac{dN}{dz}$+$d_{\rm A}$} &
\multicolumn{1}{c}{$\frac{dN}{dz}$+$d_{\rm A}$} &
\multicolumn{1}{c}{$\frac{dN}{dz}$+$d_{\rm A}$} &
\multicolumn{1}{c}{$\frac{dN}{dz}$+SNe} &
\multicolumn{1}{c}{$\frac{dN}{dz}$+CMB priors} &
\multicolumn{1}{c}{$\frac{dN}{dz}$+follow-1} &
\multicolumn{1}{c}{$\frac{dN}{dz}$+follow-2}\\
$N_{cl}/N_{SNe}$ & \multicolumn{1}{c}{995} & \multicolumn{1}{c}{578} &
\multicolumn{1}{c}{118} & \multicolumn{1}{c}{25} & \multicolumn{1}{c}{557} & \multicolumn{1}{c}{--} &
\multicolumn{1}{c}{97} & \multicolumn{1}{c}{97}\\
\multicolumn{1}{c}{} & \multicolumn{1}{c}{} & \multicolumn{1}{c}{$(N_{\rm beam}=2)$} & \multicolumn{1}{c}{$(N_{\rm beam}=3)$}
& \multicolumn{1}{c}{$(N_{\rm beam}=4)$} & \multicolumn{1}{c}{} & \multicolumn{1}{c}{on $\Omega_m$ and $\sigma_8$} 
& \multicolumn{1}{c}{$\Delta M \sim 15-50\%$} & \multicolumn{1}{c}{$\Delta M \sim 30-100\%$} \\
\hline
\hspace{0pt}$\Delta \Omega_m$ & 0.067 & 0.036 & 0.053 & 0.052 & 0.029 & 0.018 & 0.073 & 0.090\\
\hspace{0pt}$\Delta w_0$ & {\bf 0.79} & {\bf 0.36} & {\bf 0.57} & {\bf 0.63} & 0.15 & 0.40 & 0.81 & 0.80\\
\hspace{0pt}$\Delta w_a$ & {\bf 3.5} & {\bf 2.0} & {\bf 2.8} &  {\bf 2.7} & 1.0  & 2.0  & 3.4 & 3.3\\
\hspace{0pt}$\Delta \sigma_8$ & 0.12 & 0.10 & 0.10 & 0.11 & 0.10 & 0.05 & 0.06 & 0.07\\
\hline
\end{tabular}\hfil}
\label{tab:results_SPT}
\end{table*}

\begin{figure}[tb]
\begin{center}
\includegraphics[width=9cm]{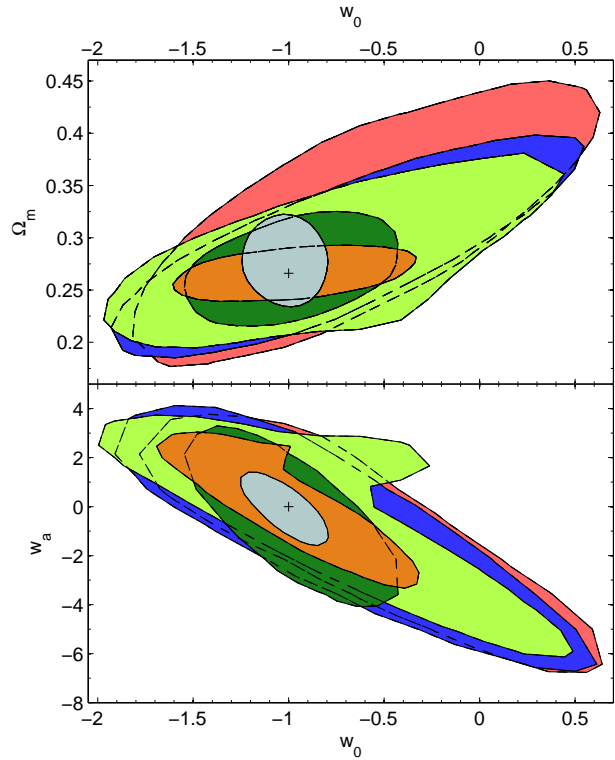}
\caption{Cosmological constraints from ACT/SPT: The 1-$\sigma$ regions in the order of overlap (decreasing area) are,
 number counts -- with mass follow-up(no priors on $A$ and $\alpha$) with
 $\Delta M \sim 30-100\%$ (red) and $\Delta M \sim 15-50\%$ (dark-blue); only number counts (light-green);
  with \da ($N_{\rm beam}=2$, 578 clusters) (dark-green);
 with CMB priors on $\Omega_m$ and $\sigma_8$ (orange); with 557 SNe from SCP Union 2 dataset (light-blue).}
\label{fig:SPT}
\end{center}
\end{figure}

\section{Results and Discussions}
\begin{figure}[tb]
\begin{center}
\includegraphics[width=9cm]{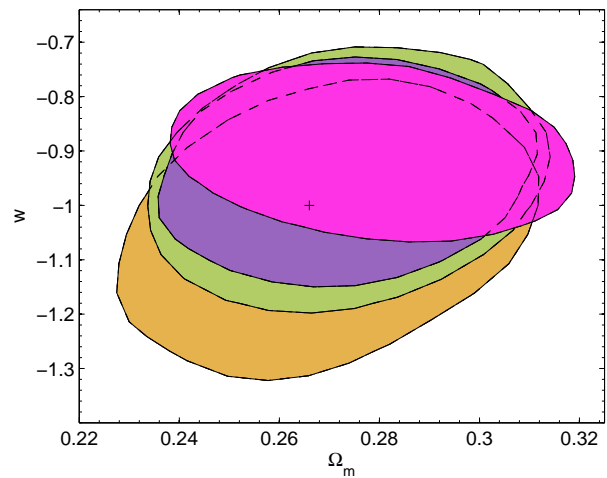}
\caption{Constraints obtained from Planck \dn added to \da measured from the overlap of Planck
plus eROSITA survey areas. The figures show 1-$\sigma$ regions in the order of overlap (decreasing area)
 as follows: just \dn (yellow); \dn with \da for $N_{\rm beam}$ = 4 (light-green), 3 (indigo) and 2 (magenta) respectively.}
\label{fig:Planck}
\end{center}
\end{figure}

Our main results are summarized in Tables \ref{tab:results_SPT} and
\ref{tab:results_Planck}, which list the 1-$\sigma$
uncertainties on the cosmological parameters. The numbers in bold highlight the improvement in the
 dark energy constraints on adding \da to \dn. The benefits of having additional
information from \da is also illustrated in figures \ref{fig:SPT} and \ref{fig:Planck}.
In all cases, we consider eROSITA as the X-Ray survey and either Planck or ACT/SPT as the SZE survey.

\begin{table}[htb]\small
\caption{Comparison of $1-\sigma$ parameter constraints from eROSITA
$\frac{dN}{dz}$ \& its overlap with Planck. \\}
\hbox to \hsize{\hfil\begin{tabular}{ccccc}
\hline
Parameter & \multicolumn{1}{c}{only $\frac{dN}{dz}$} &
\multicolumn{1}{c}{$\frac{dN}{dz}$+$d_{\rm A}$} &
\multicolumn{1}{c}{$\frac{dN}{dz}$+$d_{\rm A}$} & \multicolumn{1}{c}{$\frac{dN}{dz}$+$d_{\rm A}$}\\
 &  & \multicolumn{1}{c}{($N_{\rm beam}=2$)} &
\multicolumn{1}{c}{($N_{\rm beam}=3$)} &
\multicolumn{1}{c}{($N_{\rm beam}=4$)} \\
$N_{cl}$ & \multicolumn{1}{c}{1994} & \multicolumn{1}{c}{1829} &
\multicolumn{1}{c}{951} & \multicolumn{1}{c}{202}\\
\hline
\hspace{0pt}$\Delta \Omega_m$ & 0.027 & 0.027 & 0.026 & 0.028\\
\hspace{0pt}$\Delta w$ & {\bf 0.17}  & {\bf 0.10}  & {\bf 0.13} & {\bf 0.15} \\
\hspace{0pt}$\Delta \sigma_8$ & 0.086 & 0.099 & 0.083 & 0.093\\
\hline
\end{tabular}\hfil}
\label{tab:results_Planck}
\end{table}

\subsection{Constraints from \da added to ACT/SPT \dn}
In figure \ref{fig:SPT}, we see a significant improvement in cosmological constraints
from ACT/SPT, when $d_{\rm A}$ for $N_{\rm beam}=2$ is added to \dn. As seen in Table \ref{tab:results_SPT}
 the 1-$\sigma$ region in the dark energy plane $w_0$-$w_a$ shrinks by factors of 3.8,
 1.7 and 1.6 when the \da datasets with $N_{\rm beam}$ = 2, 3 and 4 are added to \dn;  
while in the $\Omega_m$ - $w_0$ plane it decreases by factors of 4.1, 1.8 and 1.1.
 The corresponding marginalized constraints on a single equation of state $w$ improves
 from 0.20 to 0.15 and 0.18 for $N_{\rm beam}$ = 2 and 3 respectively when \da is added to \dn.
 There is not much improvement with the smallest dataset of $N_{\rm beam}$ = 4, since the \da sample has only 25 clusters now.

Since SNe Ia are also used to measure distances ($d_{\rm L}(z)$), we compare our results with the benefits of
 adding the Union 2 compilation of the Supernova Cosmology Project (SCP) dataset in column 6. As expected,
 with the SNe data of a similar size but with much smaller errors
 (7-10\%), we get tighter constraints on dark energy with $\Delta w_0$ = 0.15 and
 $\Delta w_a$ = 1.0. Column 7 of the same table lists the effect of using CMB constraints from the WMAP7 results
 on $\Omega_m$(=0.019) and $\sigma_8$(=0.059) with the \dn data from ACT/SPT to further improve the constraints
 on dark energy with $\Delta w_0$ = 0.4 and $\Delta w_a$ = 2.0. These constraints are very similar to the ones
 through \da for $N_{\rm beam}$ = 2; compare column 3 and 7 in Table \ref{tab:results_SPT}.

Next, we compare our constraints with those that would be obtained from
dedicated follow-up observations of cluster masses requiring detailed X-Ray, SZE
or galaxy spectroscopic observations \citep{MM03}. Adding a mass follow-up to
$\frac{dN}{dz}$ constraints the cluster scaling relation and is similar to
putting priors on the scaling parameters. We build two mock catalogs
 of 97 clusters each, one with errors between 15-50\% and the other with 30-100\% errors. 
We find that the constrains from both the mass follow-up's (now without priors on the scaling
 parameters $A$ and $\alpha$) are very similar to those obtained using priors of the scaling
 relations, compare column 2 with 8 and 9 in Table \ref{tab:results_SPT}. This also implies
 that addition of \da can further improve constraints even when a mass follow-up is done. Compare
 columns 3 - 5 with 8 and 9.

\subsection{Constraints from \da added to Planck \dn}
The eROSITA survey will detect all the clusters seen by Planck and hence
together, these surveys will yield a large number of candidates for estimating  $d_{\rm A}(z)$. In our mock catalog the number of
cluster datasets for which $d_{\rm A}$ can be measured, for this combination, are found to contain 1829,
951 and 202 halos for $N_{\rm beam}$ = 2, 3 and 4, respectively. However, in contrast with ACT/SPT, Planck would be able to detect only
 the most massive clusters due to a lower resolution and mostly the ones occurring at lower redshifts. There would be a
 only very few clusters at higher redshifts ($\geq 0.6$) and hence we consider the constraints only on a constant equation of state $w$.
Table \ref{tab:results_Planck} lists the results for the improvement in the constraints from \dn when \da from an
 overlap of Planck and eROSITA are added. We see here that $\Delta w$ decreases by factors of 1.7, 1.3 and 1.1 with the addition of \da
 datasets with $N_{\rm beam}$ = 2, 3 and 4 respectively.

\section{Conclusions}
We show that just using number count observations from the ongoing SZE galaxy cluster
surveys like ACT/SPT would not be sufficient to put tight constraints on the cosmological
 parameters, especially the dark energy parameters. However these constraints can be significantly
 improved by doing a joint analysis with \da. These $d_{\rm A}(z)$ can be constructed
 out of clusters detected jointly in X-Ray and SZE
observations having overlapping areas. The optical follow-up providing the redshifts for \dn will, also,
 naturally provide the redshifts for \da. With the current and upcoming large area surveys one
 will be able to get thousands of clusters providing us with us with $d_{\rm A}$ at
 various redshifts without much effort. 

 We find that adding \da to number count observations always improves the dark energy
constraints, from \dn alone by {\it factors of 1.5 to 4}. This leads to better constraints not
 only on dark energy but also on the parameter $\Omega_m$. Even when a targeted mass follow-up of
 clusters helps in breaking the cluster-cosmology degeneracies, addition of \da helps in further
 tightening of the cosmological constraints. Moreover addition of \da improves dark energy constraints
 comparable to the improvement brought by adding CMB priors on $\sigma_8$ and $\Omega_m$. 
Thus, our proposal of adding the \da data to cluster number counts provides a natural way of
 improving the cosmological constraints using clusters alone.

The authors would like to thank Gil Holder, Jonathan Dudley and Joe Mohr for many discussions during
 the work.


\bibliographystyle{apj}

\end{document}